\documentclass[10pt]{article}
\usepackage{latexsym}
\usepackage{amssymb}
\usepackage{amsfonts}
\usepackage{amsmath}
\usepackage{theorem}

\newtheorem{theorem}{Theorem}

\newtheorem{definition}{Definition}

\begin{document}

\title{Random Multi-Overlap Structures\\
for Optimization problems}
\author{Luca De Sanctis
\footnote{Department of Mathematics,
Princeton University, Fine Hall, Washington Road,
Princeton NJ 08544--1000 \ USA \ 
{\tt<lde@math.princeton.edu>}}}

\maketitle

\begin{abstract}
We extend to the K-SAT and $p$-XOR-SAT optimization problems 
the results recently achieved, by introducing the concept
of Random Multi-Overlap Structure,  for the Viana-Bray model of diluted
mean field spin glass. More precisely one can prove a generalized bound 
and an extended variational principle for the free energy per site in the 
thermodynamic limit. Moreover a trial function implementing
ultrametric breaking of replica symmetry is exhibited.
\end{abstract}

\noindent{\em Key words and phrases:} optimization problems, 
replica symmetry breaking, ultrametric overlap structures.


\section{Introdution}

In the case of non-diluted spin glasses, M. Aizenman R. Sims and S. L. Starr (\cite{ass}) 
introduced the idea of Random Overlap Structure (ROSt) 
to express in a very elegant manner
the free energy of the model as the an infimum over a rich probability space,
to exhibit an optimal structure (the so-called Boltzmann one), to write down a
general trial function through which one can 
formulate various ansatz's for the free energy of the model.
It was also described how to formulate in particular the Parisi ansatz within this formalism.
In \cite{lds1, lds2} we extended those results to the case of diluted spin glass
(Viana-Bray model). Here we extend the same results to optimization problems,
the K-SAT and the $p$-XOR-SAT. The latter is the simple extension to $p$-body
interactions of the Viana-Bray model, which is the diluted version of the
famous model of Sherrington and Kirkpatrick of mean field spin glass.
Many of the calculations in the present paper are quite simple and 
standard, and as general reference with many details the reader can
take for instance \cite{franz}.
 

\section{Model, Notations, Definitions}

Consider configurations of Ising spins 
$\sigma:i\rightarrow\sigma_{i}=\pm1, i=1,...N$.
Let $P_\zeta$ be Poisson random variable of mean $\zeta$,
and $\{i^{\mu}_\nu\}$ be independent identically 
distributed random variables, 
uniformly distributed over points $\{1,\ldots, N\}$.
If $\{J^{\mu}_\nu\}$ are independent identically 
distributed copies of a symmetric random variable $J=\pm 1$, 
then the Hamiltonian of random K-SAT is
\begin{equation*}
H=-\sum_{\nu=1}^{P_{\alpha N}} \frac12(1+J^{1}_\nu\sigma_{i^{1}_\nu})
\cdots\frac12(1+J^{K}_\nu\sigma_{i^{K}_\nu})\ .
\end{equation*}
Here $\alpha\geq 0$ is the degree of connectivity and
both $p$ and $K$ are supposed to be even.
We do not consider the presence of an external field, but
all the results trivially extends to this case as well.
By $\omega$ we mean the Bolztmann-Gibbs 
average
\begin{equation*}
\omega(\mathcal{O})=
Z_N^{-1}
\sum_{\{\sigma\}}\mathcal{O}(\sigma)\exp(-\beta H)\ ,\  Z_N=\sum_{\{\sigma\}}
\exp(-\beta H)
\end{equation*}
We will denote by $\mathbb{E}$ 
the average over all the others (quenched) random variables,
and the free energy $f_{N}$ per site and its thermodynamic limit are defined by
\begin{equation*}
-\beta f_N=\frac1N \mathbb{E}\ln Z_N\ ,\ f=\lim_{N\rightarrow\infty}f_{N}
\end{equation*}
We will use the notation
$\Omega$ for the product of the needed 
number of independent copies (replicas) 
of $\omega$ and 
$\langle\cdot\rangle$ for the composition 
of an $\mathbb{E}$-type average over some 
quenched variables and some sort of 
Boltzmann-Gibbs average over the spin variables, 
to be specified each time.
The multi-overlaps are defined (using replicas) by
\begin{equation*}
\label{overlap}
q_{1\cdots n}=\frac{1}{N}\sum_{i=1}^N\sigma_i^{(1)}\cdots\sigma_i^{(n)}\ .  
\end{equation*}
 \begin{definition} 
 A \emph{Random Multi-Overlap Structure} 
$\mathcal{R}$ is a triple 
$(\Sigma, \{\tilde{Q}_{n}\}, \xi)$ where  
\begin{itemize}
  \item $\Sigma$ is a discrete space;
  \item $\xi: \Sigma\rightarrow\mathbb{R}_+$ 
  is a system of random weights; 
  \item 
  $\tilde{q}_{r_{1}\cdots r_{2l}}:\Sigma^{2l}\rightarrow[0, 1] ,  
  l\in\mathbb{N} , |\tilde{q}|\leq 1$ is a 
  positive definite \emph{Multi-Overlap Kernel}. 
\end{itemize}
\end{definition} 


\section{The Structure of the model}

In order to understand what is the underlying structure of the model, it is well known
that it is useful to compute the derivative of the free energy with respect to the somewhat
basic parameter. In the case of non-diluted spin glasses such parameter 
the strength of the couplings (and it is equivalent to differentiating with respect to 
the inverse temperature). In the case of diluted spin glasses such parameter is
the connectivity. \\
It is very easy to show (see e.g. \cite{franz}) by pretty standard calculation that
\begin{equation}\label{basic}
\frac{d}{d\alpha}\frac 1N\mathbb{E}\ln \sum_{\gamma}\xi_{\gamma}
\exp(-\beta H)
 =  \sum_{n>0}\frac{(-1)^{n+1}}{n}\left(\frac{e^{-\beta}-1}{2^{K}}\right)^{n}
\langle (1+Q_{n}(q))^{K}\rangle
\end{equation}
where 
\begin{equation*}
Q_{2n}(q)=\sum_{l=1}^{n}
  \sum^{1,2n}_{r_{1}<\cdots< r_{2l}}q_{r_{1}\cdots r_{2l}}\ ,\ 
  Q_{2n+1}=0\ .
\end{equation*}
The fundamental quantities governing the model are therefore the multi-overlap,
like for diluted spin glasses (\cite{lds1, lds2}). That is why we use RaMOSt
in this context as well. The main difference is that here the function $1+Q_{n}$
takes the place of the mere multi-overlaps.
 
As it should be clear from \cite{ass, lds1, lds2}, we must therefore 
introduce also two random variables 
$\tilde{H}_{.}(\gamma, \alpha; \tilde{J})$ and $\hat{H}(\gamma, \alpha; \hat{J})$ 
such that 
\begin{eqnarray*}
&&\hspace{-0.7cm}\frac{d}{d\alpha}\frac 1N\mathbb{E}\ln \sum_{\gamma}\xi_{\gamma}
\exp(-\beta\tilde{H}_{.}) 
= K\sum_{n>0}\frac{(-1)^{n+1}}{n}\left(\frac{e^{-\beta}-1}{2^{K-1}}\right)^{n}\!\!
\langle (1+Q_{n}(\tilde{q}))^{K-1}\rangle \\
&&\hspace{-0.7cm}\frac{d}{d\alpha}\frac 1N\mathbb{E}\ln \sum_{\gamma}\xi_{\gamma}
\exp(-\beta \hat{H})
= (K\!-\!1)\sum_{n>0}\frac{(-1)^{n+1}}{n}\left(\frac{e^{-\beta}-1}{2^{K}}\right)^{n}\!\!
\langle (1+Q_{n}(\tilde{q}))^{K}\rangle
\end{eqnarray*}
Finally, we introduce as expected the following trial function,
that we write this first time only considering an external field
\begin{equation*}
G_{N}(\mathcal{R}, \tilde{H}, \hat{H})=\frac 1N\mathbb{E}\ln
\frac{\sum_{\sigma, \tau}\xi_{\tau}
\exp(-\beta\sum_{i=1}^{N}(\tilde{H}_{i}+h)
\frac12(1+J_{i}\sigma_{i}))}{\sum_{\tau}\xi_{\tau}
\exp(-\beta \hat{H})}
\end{equation*}
where $\tilde{H}_{i}$ are independent copies of $\tilde{H}_{.}$
We will construct explicitly $\tilde{H}_{.}$ and $\hat{H}$ in the next sections.
Let us define
\begin{equation*}
\tilde{H}=\sum_{i=1}^{N}\tilde{H}_{i}\frac12(1+J_{i}\sigma_{i})\ .
\end{equation*}


\section{Generalized Bound and Extended Variational Principle}

Let us state the extension to the K-SAT model of the results 
presented in \cite{ass, lds1}.

Consider the interpolating Hamiltonian 
\begin{equation*}
H_{\gamma}(t)=H(t)+\tilde{H}(1-t)+\hat{H}(t)
\end{equation*}
where $t$ is understood to multiply the connectivity, and 
\begin{equation*}
R(t)=\frac{1}{N}\mathbb{E}\ln\frac{\sum_{\gamma, \sigma}
\xi_{\gamma}\exp(-\beta H_{\gamma}(t))}{\sum_{\gamma}
\xi_{\gamma}\exp(-\beta\hat{H}_{\gamma})}
\end{equation*}
then it is easy to prove (\cite{lds1, franz, ass}) by interpolation the following
\begin{theorem}[Generalized Bound]
\label{b}
\begin{equation*}
\label{ }
-\beta f\leq \lim_{N\rightarrow\infty}
\inf_{\mathcal{R}} G_N\ .
\end{equation*}
\end{theorem}
The proof is based on observing that
$R(1)=-\beta f_{N}, R(0)=G_{N}$ and computing the $t$-derivative of 
$R(t)$ using the expressions in the previous section
\begin{multline*}
\frac{d}{dt}R(t)=-\alpha\sum_{n>0}\frac{1}{2n}
\left(\frac{e^{-\beta}-1}{2^{K}}\right)^{2n}\times \\
\langle (1+Q_{2n}(q))^{K}-p(1+Q_{2n}(q))(1+Q_{2n}(\tilde{q}))^{K-1}
+(p-1)(1+Q_{2n}(\tilde{q}))^{K}\rangle
\end{multline*}
where the odd terms are missing since they cancel out. 
Therefore the derivative above is non-positive
since the function $x^{p}-pxy^{p-1}+(p-1)y^{p}$ 
of $x$ and $y$ is non-negative.

The Boltzmann RaMOSt $\mathcal{R}_{B}(M)$ is by definition the one for which
$\Sigma=\{-1,1\}^{M}$ and, using $\tau$ instead of $\gamma$, one choses 
$\xi_{\tau}=\exp(-\beta H_{M}(\tau))$ and
\begin{eqnarray*}
\tilde{H}_{\tau} &=&
-\sum_{\nu=1}^{P_{K\alpha N}}
\frac12(1+\tilde{J}^{1}_\nu\tau_{j^{1}_\nu})
\cdots\frac12(1+\tilde{J}^{K-1}_\nu\tau_{j^{K-1}_\nu})
\frac12(1+J^{K}_\nu\sigma_{i_\nu})\\
\hat{H}_{\tau} &=&
-\sum_{\nu=1}^{P_{(K-1)\alpha N}}
\frac12(1+\hat{J}^{1}_\nu\tau_{j^{1}_\nu})
\cdots\frac12(1+\hat{J}^{K}_\nu\tau_{j^{K}_\nu})
\end{eqnarray*}
where the independent random variables $j_{.}^{.}$ are
uniformly distributed over $1,\ldots,M$
and $\tilde{J}_{.}^{.}, \hat{J}_{.}^{.}$ are independent copies of $J$.
The (limiting) Boltzmann RaMOSt $\mathcal{R}_{B}$
fulfills the Reversed Bound (\cite{ass, lds1})
\begin{equation*}
-\beta f\geq\lim_{N\rightarrow\infty}\liminf_{M\rightarrow\infty}
G_{N}(\mathcal{R}_{B}(M))=\lim_{N\rightarrow\infty}G_{N}(\mathcal{R}_{B})
\end{equation*}
which can be proven in the same way as the analogous 
theorem 3 of \cite{lds1}, except here we must chose
$\alpha^{\prime}=\alpha(1+(K-1)N/M)$.
As a consequence, one can state the following
\begin{theorem}[Extended Variational Principle]
\begin{equation*}
\label{evp}
-\beta f = \lim_{N\rightarrow\infty}\inf_{\mathcal{R}}G_N\ .
\end{equation*}
\end{theorem}
 It is easy to see that the Boltzmann RaMOSt for the K-SAT is factorized in the sense
 of section 5 of \cite{lds1}.
 

\section{Replica Symmetry Breaking and Ultrametric RaMOSt}\label{rsb}

We are about to extend the results of \cite{lds2} to the K-SAT by constructing the 
Ultrametric RaMOSt $\mathcal{R}_{U}$ with $R$-level Replica Symmetry Breaking.
The latter corresponds to the choice of 
$\xi_{\gamma}(m_{1}, \ldots , m_{R}), \gamma=(\gamma_{1}, \ldots , \gamma_{R})$ 
(illustrated e.g. in \cite{t1}) derived from the Random Probability Cascades, to be used 
in the case of the K-SAT together with
\begin{eqnarray*}
\tilde{H}_{\gamma} &=&
\sum_{\nu=1}^{P_{K\alpha N}}
\tilde{u}_{\nu}^{\gamma}
\frac12(1+J_\nu\sigma_{i_\nu})
-\frac{1}{\beta}\ln\cosh(\beta \tilde{u}^{\gamma}_{\nu})\\
\hat{H}_{\gamma} &=&
\sum_{\nu=1}^{P_{(K-1)\alpha N}}
\hat{u}_{\nu}^{\gamma}
-\frac{1}{\beta}\ln\cosh(\beta \hat{u}^{\gamma}_{\nu})
\end{eqnarray*}
with $\tilde{u}_{\gamma}, \hat{u}_{\gamma}$ chosen such that
\begin{eqnarray*}
\tanh(\beta\tilde{u}_{\gamma}) &=&
(e^{-\beta}-1)\frac12(1+\tilde{J}^{1} W^{1}_{\gamma})
\cdots\frac12(1+\tilde{J}^{K-1}W^{K-1}_{\gamma})\\
\tanh(\beta\hat{u}_{\gamma}) &=&
(e^{-\beta}-1)\frac12(1+\hat{J}^{1} W^{1}_{\gamma})
\cdots\frac12(1+\hat{J}^{K}W^{K}_{\gamma})
\end{eqnarray*}
in which $W_{\gamma}$ is the same as in the case of diluted $p$-spin glasses
(\cite{lds2})
\begin{equation*}
W_{\gamma}=\tilde{\omega}_{\tilde{\alpha}_{1}}
(\rho_{k_{\nu}})\bar{J}_{\gamma_{1}}+\cdots+
\tilde{\omega}_{\tilde{\alpha}_{R}}
(\rho_{k_{\nu}})\bar{J}_{\gamma_{1}\cdots\gamma_{R}}
\end{equation*}
where $\tilde{\omega}_{\tilde{\alpha}}(\rho_{k_{\nu}})$ is the infinite volume limit
of the Boltzmann-Gibbs 
average of a random spin from an auxiliary system with a
Viana-Bray one-body interaction Hamiltonian
at connectivity $\tilde{\alpha}$ (\cite{lds2}).
The indices, the bar, the tilde, the hat mean 
independent copies of the corresponding variables.
Let us now report a comment from \cite{lds1}.
Given any partition $\{x^{a}\}_{a=0}^{R}$ of 
the interval $[0 , 1]$, there exists a sequence 
$\{\tilde{\alpha}_{a}\}_{a=0}^{R}\in[0 , \infty]$ such that
$\tilde{q}_{1\cdots n}(\tilde{\alpha}_{a})=x_{a}-x_{a-1}$. In other words,
a sequence $\{\tilde{\alpha}_{a}\}_{a=0}^{R}\in[0 , \infty]$ generates
for each $n\in \mathbb{N}$ a partition of $[0 , 1]$ considered as 
the set of trial values of $\tilde{q}_{1\cdots n}$, provided the $\tilde{\alpha}_{a}$
are not too large
\begin{equation}\label{constraint}
\sum_{a\leq R}\tilde{q}_{1\cdots n}(\tilde{\alpha}_{a})\leq 1\ .
\end{equation}
 We limit our trial 
multi-overlaps to belong to partions generated in this way.
This implies that the points of the generated partitions tend to
get closer to zero as $n$ increases. Which is good, since
in any probability space $\langle \tilde{q}_{n}\rangle$ decreases
as $n$ increases and therefore the probability integral 
distribution functions tend to grow faster near zero.
Now put inductively
\begin{equation*}
\mathbb{E}\tilde{\Omega}_{\tilde{\alpha}_{a}}
(\tau^{(r_{1})}_{k_{\nu}}\cdots\tau^{(r_{l})}_{k_{\nu}})=
\tilde{q}_{r_{1}\cdots r_{l}}(\tilde{\alpha}_{a})=
\tilde{q}^{(a)}_{r_{1}\cdots r_{l}}-
\tilde{q}^{(a-1)}_{r_{1}\cdots r_{l}}\ ,\ \tilde{q}^{(0)}_{r_{1}\cdots r_{l}}=0
\end{equation*}
then an elementary calculation shows that
\begin{eqnarray*}
\mathbb{E}\tanh^{n}(\beta\tilde{u}_{\gamma})&=&
\left(\frac{e^{-\beta}-1}{2^{K-1}}\right)^{n}(1+Q_{n}(\tilde{q}))^{K-1}\\
\mathbb{E}\tanh^{n}(\beta\hat{u}_{\gamma})&=&
\left(\frac{e^{-\beta}-1}{2^{K}}\right)^{n}(1+Q_{n}(\tilde{q}))^{K}
\end{eqnarray*}
with $\tilde{q}$ ultrametric (\cite{lds2}), i.e.
\begin{equation*}
\tilde{q}_{r_{1}\cdots r_{l}}=(\tilde{q}^{(1)}_{r_{1}\cdots r_{l}}-
\tilde{q}^{(0)}_{r_{1}\cdots r_{l}})\delta_{\gamma^{r_{1}}_{1}
\cdots\gamma^{r_{l}}_{1}}
+\cdots+(\tilde{q}^{(R)}_{r_{1}\cdots r_{l}}-\tilde{q}^{(R-1)}_{r_{1}\cdots r_{l}}) 
\delta_{\gamma^{r_{1}}_{1}\cdots\gamma^{r_{l}}_{1}} \cdots\delta_{\gamma^{r_{1}}_{R}\cdots\gamma^{r_{l}}_{R}}
\end{equation*}
If we denote by $X$ the map $\tilde{\alpha}_{a}\rightarrow m_{a}$ satisfying 
(\ref{constraint}), we 
have $\mathcal{R}_{U}=\mathcal{R}_{X, R}$ and the trial function is
\begin{equation*}
\inf_{X, R}G(\mathcal{R}_{X, R})\ .
\end{equation*}
Notice that the Ultrametric RaMOSt for the K-SAT factorizes 
exactly like the Boltzmann one, which is optimal.


\section{Conclusions}

The RaMOSt is the minimal generalization of the ROSt, and what we showed here 
and in \cite{lds1} is 
that the minimal generalization is enough to formulate the variational principle and also
exhibit a concrete RaMOSt analogous to the Parisi one for SK. As a consequence,
it is enough to restrict the space of trial functions to those expressible in terms of fixed
multi-overlaps
(i.e. a set of numbers, not random variables to be averaged).


\appendix

\section{The $p$-XOR-SAT}\label{p-spin}

The Hamiltonian of the random $p$-XOR-SAT 
coincides with the one of the diluted $p$-spin glass 
\begin{equation*}
H=-\sum_{\nu=1}^{P_{\alpha N}} J_\nu \sigma_{i^{1}_\nu}\cdots\sigma_{i^{p}_\nu}
\end{equation*}
It is therefore elementary to extend all the results of \cite{lds1, lds2} to this case,
also when in presence of an external field.
Since it is easy to show (\cite{lds1})
\begin{equation*}
\frac{d}{d\alpha}\frac 1N\mathbb{E}\ln \sum_{\gamma}\xi_{\gamma}
\exp(-\beta H)
 =  \sum_{n>0}\frac{1}{2n}\mathbb{E}\tanh^{2n}(\beta J)
(1-\langle q^{p}_{2n}\rangle)
\end{equation*}
the structure of the model is the same RaMOSt valid for the case of the Viana-Bray model,
but the equality above suggests to try and get the non-negative convex
function $x^{p}-pxy^{p-1}+(p-1)y^{p}$ whenever we got the square $x^{2}-2xy+y^{2}$
in the Viana-Bray  case.\\
That is why here $\tilde{H}$ and $\hat{H}$ are chosen such that
\begin{eqnarray*}
\frac{d}{d\alpha}\frac 1N\mathbb{E}\ln \sum_{\gamma}\xi_{\gamma}
\exp(-\beta\tilde{H}_{.})
& = & p\sum_{n>0}\frac{1}{2n}\mathbb{E}\tanh^{2n}(\beta J)
(1-\langle \tilde{q}^{p-1}_{2n}\rangle)\label{eta} \\
\frac{d}{d\alpha}\frac 1N\mathbb{E}\ln \sum_{\gamma}\xi_{\gamma}
\exp(-\beta \hat{H})
&=& (p-1)\sum_{n>0}\frac{1}{2n}\mathbb{E}\tanh^{2n}(\beta J)
(1-\langle\tilde{q}^{p}_{2n}\rangle)\label{kappa}
\end{eqnarray*}
and plugged in
\begin{equation*}
G_{N}(\mathcal{R}, \tilde{H}, \hat{H})=\frac 1N\mathbb{E}\ln
\frac{\sum_{\sigma, \tau}\xi_{\tau}
\exp(-\beta\sum_{i=1}^{N}(\tilde{H}_{i}+h)\sigma_{i})}{\sum_{\tau}\xi_{\tau}
\exp(-\beta \hat{H})}
\end{equation*}
The Generalized Bound clearly holds, 
the Boltzmann RaMOSt is the one with  
\begin{equation*}
\tilde{H}_{\tau} =
-\sum_{\nu=1}^{P_{p\alpha N}}
\tilde{J}_{\nu}\tau_{j^{1}_{\nu}}\cdots\tau_{j^{p-1}_{\nu}}\sigma_{i_{\nu}}\ ,\
\hat{H}_{\tau} =
-\sum_{\nu=1}^{P_{(p-1)\alpha N}}
\hat{J}_{\nu}\tau_{j^{1}_{\nu}}\cdots\tau_{j^{p}_{\nu}}\ ,\ 
\xi_{\tau}=e^{-\beta H_{M}}
\end{equation*}
(same couplings as the original system)
and it is optimal so that we can also state the Extended Variational Principle.
The Broken Replica Symmetry Ultrametric RaMOSt (which includes as a trivial case the Replica Symmetric one) relies on the weights $\xi_{\gamma}$ of the Random Probability Cascades
as in \cite{t1} and on
\begin{eqnarray*}
\tilde{H}_{\gamma} &=&
-\sum_{\nu=1}^{P_{p\alpha N}}\left(\frac1\beta
\ln\frac{\cosh(\beta J)}{\cosh(\beta 
\tilde{J}^{\gamma}_{\nu})}
+\tilde{J}^{\gamma}_{\nu}\sigma_{i_{\nu}}\right)\\
\hat{H}_{\gamma} &=&
-\sum_{\nu=1}^{P_{(p-1)\alpha N}}\left(\frac1\beta
\ln\frac{\cosh(\beta J)}{\cosh(\beta 
\hat{J}^{\gamma}_{\nu})}
+\hat{J}^{\gamma}_{\nu}\right)
\end{eqnarray*}
with
\begin{equation*}
\label{ }
\tanh(\beta\tilde{J}_{\gamma})=
\tanh(\beta J)\tilde{W}^{1}_{\gamma}\cdots\tilde{W}^{p-1}_{\gamma}\ ,\ 
\tanh(\beta\hat{J}_{\gamma})=
\tanh(\beta J)\tilde{W}^{1}_{\gamma}\cdots\tilde{W}^{p}_{\gamma}
\end{equation*}
where $\tilde{W}^{.}_{\gamma}$ are independet copies of 
\begin{equation*}
\label{ }
\tilde{W}_{\gamma}(\bar{J}, k_{\nu})=
\tilde{\omega}_{\tilde{\alpha}_{1}}(\rho_{k_{\nu}})\bar{J}_{\gamma_{1}}+\cdots+
\tilde{\omega}_{\tilde{\alpha}_{R}}(\rho_{k_{\nu}})\bar{J}_{\gamma_{1}\cdots\gamma_{R}}
\end{equation*}
with $\bar{J}_{.}=\pm 1$ symmetric.


\section*{Acknoledgments}

The author warmly thanks Francesco Guerra for useful 
discussions, and gratefully 
acknowledges the hospitality of the Department of 
Physics at University of Rome
``La Sapienza'' (and in particular 
Giovanni Jona-Lasinio). 


\end{document}